\documentstyle [12pt] {article}
\def\ni{\noindent}

\def\deg{\ifmmode^\circ\else$^\circ$\fi}
\def\solar{\ifmmode_{\mathord\odot}\else$_{\mathord\odot}$\fi}

\def\nott{\ifmmode_\circ\else$_\circ$\fi}

\title{ \Large \bf Magnetic Activity in Thick Accretion
Disks and Associated Observable Phenomena I. Flux Expulsion}
\author
{Sandip K. Chakrabarti and Sydney D'Silva\\
Tata Institute of Fundamental Research, Homi Bhabha Road, Colaba,\\
Bombay, 400005, INDIA\\}
\begin{document}
\baselineskip 22pt
\maketitle
\begin{abstract}

We study the dynamics of toroidal magnetic flux tubes, symmetric
about the rotation axis, inside non-magnetic thick accretion disks
around black holes. We present model equations which include
effects of gravity, centrifugal force, pressure gradient force,
Coriolis force, drag, magnetic tension and magnetic buoyancy. We
solve them assuming the disk to be adiabatic. We show that under a
wide range of parameters describing the size and the field strength,
as well as angular momentum distribution inside the disk,
buoyant flux tubes, either released on the equatorial plane or at the
outer edge of the disk, can gather in the chimney-like openings near
the axis. This behavior makes
the chimneys magnetically most active and could shed light on the origin and
acceleration of cosmic jets, as well as the variabilities observed in Blazars.

\end{abstract}


\newpage

\noindent {\large \bf{ 1. INTRODUCTION}}

Understanding the formation, acceleration and collimation of radio jets
is one of the major challenges faced by astrophysicists.
Several models (e.g., Begelman, Blandford \& Rees, 1984 for a review)
invoke radiation pressure supported (Paczy\'nski \& Wiita 1980) or ion
pressure supported (Rees et al. 1982) thick accretion disks around the
central black holes. Near the axis, there is a vortex-like opening
on either side of the disk (Lynden-Bell 1978),
which is formed due to strong centrifugal force close to the axis. It is
assumed that matter continuously ejected from this opening
in the inner part of the disk (which we call `chimney' for short)
is collimated and accelerated by hydrodynamic (Fukue 1982;
Chakrabarti 1986) or hydromagnetic processes (Blandford \& Payne 1982;
Chakrabarti \& Bhaskaran 1992). Another unresolved issue of equal importance
is to understand the cause of the variabilities observed in Blazars.
Variability on timescales significantly less than a
day have now been conclusively seen in the optical band for Blazars
(Miller 1988; Miller, Carini, \& Goodrich 1989; Wagner et
al.\ 1990; Carini et al.\ 1992; Quirrenbach et al.\ 1991).
Similar rapid changes have long been known in the X-ray emission from AGN
(Treves et al.\ 1982; Lawrence et al.\ 1987; McHardy \& Czerny 1987;
Abraham \& McHardy 1989). Clear microvariability in the UV for PKS 2155-304
(Edelson et al.\ 1991) as well as intraday variations in the
radio band (Quirrenbach et al.\ 1989a,b, 1991; Krichbaum,
Quirrenbach, \& Witzel 1992) are also remarkable. Some
quasi-periodicities of $\sim 1$ and $\sim 7$ days were found for the BL Lac
0716+714 in both the radio and optical (Wagner et al.\ 1990;
Quirrenbach et al.\ 1991) bands.

No solid physical understanding of this microvariability has yet
emerged. A plausible model involves the relativistic shock-in-jet models
(Blandford \& K\"onigl 1979; Scheuer \& Readhead 1979; Hughes, Aller, \&
Aller 1985; Marscher \& Gear 1985). These models require magnetic
activity at the base of the jet. Magnetic fields are believed to be
steadily brought inside the accretion disk by the inflow. They are compressed
by the converging disk, amplified by the shear, convection and advection
and finally eliminated from the disk by the buoyancy effects (e.g.,
Chakrabarti 1990). Qualitative studies of the effects of shear
and buoyancy on these magnetic flux tubes in the context of thin
accretion disks have been made in the literature
(Eardley \& Lightman 1975; Galeev, Rosner \& Vaiana 1979; Coroniti 1981;
Shibata, Tajima \& Matsumoto, 1990).
It is observed that {\it if} the buoyancy time-scales are larger
than the shear amplification time-scale, flux tubes can be exponentially
amplified within dynamical time, until they reach equipartition
value (with local gas pressure matching magnetic pressure).
It is suggested that flux tubes, in general, rise rapidly outside
the thin disk and form a coronal structure on the disk surface
which may be responsible for
X-ray emission from the disk. Actual study of the dynamics of flux tubes
in the context of thin accretion disks has been done by
Sakimoto and Coroniti (1989), where the emphasis lies in a
consistency-check of the $\alpha$-disk hypothesis in thin disks.
A recent review on the dynamics of slender flux tubes in thin
accretion disk could be found in Schramkowski \& Achterberg (1993).

So far, no study of the flux tube behavior and its possible
influence on the jet formation and microvariability in the context of
thick accretion disks has been performed. In thick disks, the angular momentum
close to the central engine, typically assumed to be a black hole,
is almost constant (Fishbone \& Moncrief 1976; Lynden-Bell 1978; Kozlowski,
Jaroszynski \& Abramowicz, 1978). This behavior has been shown to be
due to strong radiation pressure (Maraschi, Raina \& Trevis 1976).
A thick disk has the resemblance of a toroidal star, with the center
having maximum temperature and pressure. Both the temperature and the pressure
falls off very rapidly toward the axis of the black hole and rather gradually
in the direction away from it. Magnetic flux tubes brought near the
black hole by accreting matter are expected to be buoyant
in the direction of the pressure gradient force and are
expected to move, not vertically, but toward the direction of the local
pressure gradient; although other effects, such as Coriolis, drag and
magnetic tension forces and shear amplification coupled with reconnection
of flux tubes will certainly control the detailed behavior.

In the present series of papers, we study how and where magnetic flux
tubes emerge from a thick disk and whether a thick disk can support
a coronal structure. The motivation for this work stems
from the fact that although the inner part of the chimney
(i.e., the funnel) has been long known to be the
most luminous region of the entire disk (e.g., Madau 1988),
it has not at all been demonstrated that the chimney itself is also by far
magnetically the most active region. We prove that the chimney could be very
active, and claim that the microvariabilities observed in the Blazars and AGNs
as well as observed high energy gamma rays could be due to this magnetic
activity. {\it In the present paper}, we investigate the  circumstances  under
which flux tubes released at the equatorial plane or at the outer edge of the
disk would appear in the chimney of the disk due to the well known Parker
(1955) instability and other effects. In the absence of
Coriolis force on the flux tubes in the local rotating frame, (which is
possible if the angular momentum is constant throughout the disk)
the buoyant flux tubes rise along the pressure gradient force,
bringing only a fraction of the flux tubes into the chimney.
When the angular momentum varies, Coriolis force is also felt by the moving
tubes and it plays a major role in shaping the dynamics of the flux tube.
It makes the flux tubes emerge more or less parallel
to the rotational axis, thus increasing the
fraction of tubes that enter the chimney. In the presence of
an accreting flow, this fraction is found to increase further,
bringing in flux tubes even from the outermost edge of the disk.

Magnetic fields can either be `brought in' to the accretion disk advectively
or generated by a dynamo process inside the disk from some seed field
(Eardley \& Lightman 1975; Galeev, Rosner \& Vaiana 1979; Parker 1979, Chapter
22; Soward 1978; Meyer \& Meyer-Hofmeister 1983; Pudritz 1981a, b; Stepinski \&
Levy 1988, 1990; Vishniac, Jin \& Diamond 1990; Campbell 1992; Tout \&
Pringle 1992). We show that there is always a large parameter space spanned by
flux tube size and the internal field strength in which flux tubes can
be pushed into the chimney by the Parker instability (1955), thus making the
chimney magnetically very active. Magnetic fields could also be produced
{\it ab initio} through a Bierman battery
effect (Bierman 1950) as inside proto-galactic tori (Chakrabarti 1991).
If latter is the case, the field amplified inside to local equipartition
value, must be expelled away very efficiently in order that the intergalactic
magnetic fields $\leq 10^{-9}$G be accounted for (Chakrabarti 1991).
One of the goals of the present series is to show that such expulsion is
indeed possible.

Sustained magnetic activity on the surface of the sun depends on the
fact that fields are anchored.  {\it In the next paper} of this series
(D'Silva and Chakrabarti 1993, hereafter Paper II), we show how the flux
tubes can be anchored inside the accretion disk so
as to produce a stable corona (Galeev, Rosner \& Vaiana 1979 and Stella \&
Rosner 1984 for thin disks) in the chimney. Using simple but sufficiently
reasonable equation of state we show that our disks can have zones akin to
the radiative and convective zones of a stellar interior.
Thus, these disks could, in principle, have coronae.
We discuss a few observational effects which may be produced by such
a coronal structure.

In order to understand the dynamics of the flux tubes better,
we use a very simple model of the accretion disk as well as
that of the flux tube. We use a power law angular momentum
distribution inside the disk around a Schwarzschild black hole (Paczy\'nski
\& Wiita 1980; Chakrabarti 1985). The geometry of the spacetime external
to the black hole is mimicked by using a
pseudo-Newtonian potential prescription of Paczy\'nski and Wiita (1980).
A more general potential, which we do not use here, mimicks the Kerr black
holes geometry (Chakrabarti and Khanna 1992) may be used to the study
the effects of dragging of frames also.  The flux tubes are assumed to be
axisymmetric about the rotation axis of the disk and contain only the
azimuthal component of the field as shown in Figure 1a.
The inflow velocity is added using an $\alpha$ viscosity
(Shakura \& Sunyaev, 1973) model.

In the next Section, we
present equations governing the structure of a thick disk and the
equations of motion of the flux tubes. We also study very briefly the
character of the solutions analytically. In \S 3, we present the
principal numerical results using the adiabatic
accretion disk. In the final Section, we briefly summarize our results.

\noindent {\large {\bf 2. MODEL EQUATIONS }}

We divide this Section into several sub-sections.
The first sub-section provides the equations which describe the
thick accretion disk around a black hole. We prescribe thermodynamic quantities
inside an adiabatic disk. In \S 2.2, we present
the basic equations of motion of a flux ring symmetric around the rotation
axis. In \S 2.3, we give the equations that describe how quantities
{\it inside} a flux tube evolve.
Two possibilities are distinguished: (i) flux tubes that are locally
isothermal with its surroundings and (ii) flux tubes that move adiabatically.
In \S 2.4, we present approximate analytical solutions to gain some
insight into the behavior of the flux tubes.

Throughout the paper, we shall use the spherical polar coordinates:
$r, \ \theta , \phi $ and the geometric units. Masses are
measured in units of the mass of the central black hole, $M_{BH}$; distances
will be measured in units of the Schwarzschild radius $r_g=2GM_{BH}/c^2$;
and the time scales are measured in units of $r_g/c$.

\noindent{\bf 2.1 Equations Describing a Thick Accretion Disk}

The force balance condition which determines the structure of a
thick accretion disk is given by,
\begin{equation}
\frac{{\bf \nabla} P}{\rho} = -{\bf \nabla} \phi +  \frac{l^2}{(r\sin\theta)^3}
{\bf  \nabla} R
\label{force-balance}
\end{equation}
where, $P$ and $\rho$ are isotropic pressure and density respectively,
$\phi=-GM_{BH}/(r-r_g)$ is the pseudo-Newtonian gravitational potential
of a black hole (Paczy\'nski \& Wiita 1980), $l$ is the specific angular
momentum inside the disk and $R=r \sin\theta$ is the axial distance.
In the absence of a complete theory of viscous
flow around a black hole, we assume that the angular momentum distribution
is a power law $l(\xi, \theta)=
l\nott (\xi\sin\theta )^n$, inside the disk ($\xi$
is the dimensionless radial coordinate and $\theta$ is the angle
measured from the $z$ axis). Here,
$l\nott$ and $n$ are positive constants, with $0.5 \geq n \geq 0$.
The positivity of $n$ is required by the stability criterion under
axisymmetric perturbation $dl/d\xi \geq 0$. Our choice of angular momentum
distribution is similar
to that assumed by Paczy\'nski \& Wiita (1980) and is the
reminiscent of the `natural' distribution used in any stationary, axisymmetric
spacetime (Chakrabarti 1985). The constant $l\nott$ is determined from the
consideration that at the inner edge, $l \leq l_{mb}$, where, $l_{mb} =2$
is the marginally bound angular momentum. For simplicity, we shall assume that
the equation of state is barotropic, i.e., $P=P(\rho)$. In this case, the left
hand side of the equation becomes integrable $W(P)=-\int{{dP}/{\rho}}$.
Here, $W(P)=${\it constant} surface represents an equipotential surface,
which is also a surface of constant pressure, density and temperature.
These surfaces are obtained by integrating equation (1) as,
\begin{equation}
W(P)-W(P_{out})=-\frac{1}{2(\xi-1)}-\frac{l_{out}^2}{2n-2}(\xi
\sin\theta)^{2n-2},
\label{eq:energy}
\end{equation}
\ni where $W(P_{out})$ is the integration constant signifying the
potential at the outermost surface with pressure $P_{out}$. For a
given pair ($l\nott,\ n$) there exists a point inside the disk
where the potential is minimum which corresponds to the center
of the disk. As in the case of stars, the potential gradually
rises outwards till it reaches $W(P_{out})$.

Inside the disk, we choose the polytropic equation of state,
$P=K\rho^\gamma$, where $K$ and $\gamma$ are constants. When the accretion
rate is high compared to the Eddington rate, photon and matter act as a single
fluid of polytropic index $\gamma=4/3$ and the radiation pressure
dominated geometrically thick disks are also optically thick due to Thompson
scattering (Paczy\'nski 1982). Using the above equation of state, one
obtains the expression for temperature (in \deg K) to be
\begin{equation}
\label{eq:addisk}
T_e(\xi) = T_e(R_{out}) - ({\mu\beta c^2\over{4R_G}}) \{{1\over
2}[{1\over{\xi_{out}-1}} - {1\over{\xi - 1}}] +
{l\nott^2\over{2n-2}}[\xi_{out}^{2n-2} - (\xi\sin\theta)^{2n-2}]\},
\end{equation}
where $R_{out}=\xi_{out} r_g$ is the distance of the outer
edge of the disk on the equatorial plane and $R_G$, the gas constant.
We have used the subscript $e$ to indicate that these
values are `external' to the moving flux tubes. Temperature is maximum
at the center of the disk and gradually falls off as one moves
away from it. The variation of quantities
internal to the flux tubes would be discussed in \S 2.3.
The constant $\beta= p_g/p_t$ represents the ratio of the gas pressure
to total pressure $p_t=p_g+p_r$, $p_r$ being the radiation pressure
given by $p_r=\frac{1}{3}aT_e^4$ in local thermodynamic equilibrium,
with $a$ as the radiation density constant.
The adiabatic disk is taken to be radiation pressure dominated and
the value of $\beta$ is fixed at $10^{-4}$, unless specified otherwise.
Detailed thermodynamic conditions inside the disk with this equation of state
as well as calculation of velocities in `slim' disk approximation
is discussed in Chakrabarti, Jin and Arnett (1987).
For the purpose of the present paper, we assume that the
gas is homogeneous inside and $\mu=0.5$ to be the the mean molecular weight
of the gas, which is valid for a fully ionized hydrogen.

\noindent{\bf 2.2 Equation of Motion of the Flux tubes inside the Thick Disk}

The behavior of flux tubes has been extensively studied inside
the sun. In the last four decades, magnetic activity on the surface of the
sun has attracted a great deal of attention. The most obvious question
which is asked is: how and where
do these magnetic fields come from? The pioneering work of Parker (1955) showed
that magnetic buoyancy is decidedly important in bringing out internally
generated flux tubes  on the sun. Since then a significant amount of work has
been completed, some studying the stability
of a continuous distribution of magnetic field (Gilman 1970, Acheson 1978),
while some others (Spruit \& van Ballegooijen 1982; van Ballegooijen 1982;
1983; Choudhuri \& van Ballegooijen 1988; Choudhuri \& Gilman 1987;
Choudhuri 1990; Choudhuri \& D'Silva 1990; Moreno-Insertis, Sch\"ussler,
\& Feriz-Mas 1992) studying the dynamics of the
flux tubes in a non-magnetic environment. These works essentially include
effects such as rotation, drag, etc. on the behavior of the flux tubes
and provide a better understanding of the magnetic activities which
are observed on the solar surface.

In Section 2.1 above, we have shown that thick accretion disks have
properties similar to stars except for the topological property; stars
are spherical whereas thick accretion disks are toroidal. However,
there are several ways in which disks are different. First of all,
the effect of rotation is far more important inside the disk, than
inside the sun, since matter with significant angular momentum comes
much closer to the axis. Secondly, unlike solar or stellar case,
disks have a very strong differential rotation inside.
Thirdly, flux tubes expelled by the disk are not necessarily produced
within; they could be brought in from outside, along with the accreted
material. The size as well as strength of the flux tubes could be
totally random. Each of the flux tubes could be repeatedly sheared, advected,
convected and reconnected. Under these conditions studying the dynamics
of flux tubes thoroughly would be prohibitive unless some simplifying
assumptions are made regarding their nature. We use the thin flux tube
approximation, which assumes that variations of different physical
quantities within the flux tube are small, and consider only axisymmetric
flux tubes which are symmetric about the rotational axis of the disk.
Considering that shear is very strong in an accretion disk, the latter
assumption is well justified. The
thin flux tube approximation holds good provided the flux tube
cross-sectional radius is smaller than the local pressure scale heights
of the disk. Figure 1b shows the pressure contours of a
typical adiabatic thick accretion disk with $l\nott = 2$, $n=0$, $R\nott=20
r_g$
and mass of the black hole $= 10^6$ M\solar. The contours are at the
e-folded values and plotted using Eqn. (\ref{eq:addisk}). Note that
flux tubes smaller than $0.1 r_g$ are perfectly valid in this formalism,
except in the very outermost edge of the disk where the pressure drops
off very rapidly. Similarly, while studying advection, we consider only a
small radial motion: $v_r << v_\phi$, and $v_\theta=0$. These assumptions
enable us to de-couple various physical processes affecting the behavior
of the flux tubes while retaining the salient features.

The equation of motion for thin flux tubes has been written by several
authors in the context of solar physics (See Choudhuri and Gilman, 1987;
Choudhuri 1990, and references within). For the purpose of the present work,
we use (21-23) of Choudhuri (1990) modified suitably to include the
disk effects mentioned above. The additional features included are:
(a) The acceleration due to gravity is replaced by its
effective value $g\hat{\bf
r}-{\bf\bigtriangledown}(|{\bf\Omega}\times {\bf r}|)^2$, which
includes the centrifugal force term (in the case of the sun, the
centrifugal force term is very small compared to $g$ and hence
is neglected), (b) A drag term ${\bf D}$, and (c)
the additional term in the Lagrangian derivative of $\Omega (r)$,
${\bf v}.{\bf\bigtriangledown}\Omega$ for the non-rigidity of the disk.

The equations of motion of a slender, axisymmetric flux tube which we use
in our present analysis are given by:

\begin{eqnarray}
(m_i + m_e)[{d^{2}r\over dt^{2}} - r({d\theta\over dt})^2] +
m_i[-r({d\phi\over dt})^2\sin^2{\theta} - 2r\Omega({d\phi\over
dt})\sin^2{\theta}] =\nonumber\\
 = (m_e - m_i)[g - r\Omega^2\sin^2\theta] - {\Psi^2\over
2\pi\sigma^2}\sin\theta + D_r2\pi r\sin\theta,
\label{eq:ARC1}
\end{eqnarray}

\begin{eqnarray}
(m_i + m_e)[r{d^2\theta\over dt^2} + 2{dr\over
dt}{d\theta\over dt}] + m_i[ -r({d\phi\over
dt})^2\sin\theta\cos\theta - 2r\Omega({d\phi\over
dt})\sin\theta\cos\theta] =\nonumber\\
 = -(m_e - m_i)r\Omega^2\sin\theta\cos\theta -{\Psi^2\over
2\pi\sigma^2}\cos\theta + D_\theta 2\pi r\sin\theta,
\label{eq:ARC2}
\end{eqnarray}

\begin{eqnarray}
 m_i[r{d^2\phi\over dt^2}\sin\theta + 2{dr\over
dt}{d\phi\over dt}\sin\theta + 2r{d\theta\over dt}{d\phi\over
dt}\cos\theta + 2\Omega(r{d\theta\over dt}\cos\theta + {dr\over
dt}\sin\theta)] + \nonumber\\
\nopagebreak + r\sin\theta[\dot{r}{\partial\Omega\over\partial
r} + \dot{\theta}{\partial\Omega\over\partial\theta}] = D_\phi
2\pi r\sin\theta.
\label{eq:ARC3}
\end{eqnarray}

\ni Here $(r,\theta,\phi)$ is the position of a fluid particle inside the
flux ring with magnetic field $B$, radius of cross-section
$\sigma$ and mass $m_i = \rho_i\pi \sigma^2.2\pi r\sin\theta$,
while $m_e = \rho_e\pi \sigma^2.2\pi r\sin\theta$ is the mass of
the external fluid displaced by the flux ring. Subscripts $e$
and $i$ indicate whether the relevant quantity is of the ambient
(external) medium or within the flux tube (internal medium). The
time derivatives are Lagrangian derivatives. The flux through
the tube is $\Psi = B\pi\sigma^2$.

According to the chosen $l$ distribution, angular velocity takes the form
$$\Omega = l\nott({r\over r_g}\sin\theta)^{n}{r_gc\over
(r\sin\theta)^2}.$$  The $\phi$ component of the equation of motion for a
flux ring in the presence of differential rotation is given in Appendix B
of Choudhuri and Gilman (1987) as

$${d\over dt}[r^2\{\Omega(r,\theta) + ({d\phi\over
dt})\}\sin^2\theta] = 0,$$

\ni where the Lagrangian derivative of $\Omega$ is

$${d\Omega\over dt} = {\partial\Omega\over\partial t} + {\bf
v}.{\bf\bigtriangledown}\Omega,$$

\ni which in the present case becomes

$${d\Omega\over dt} =
r_gc{l\nott\over r_g^n} (n-2)(r\sin\theta)^{(n-3)}[\dot{r}\sin\theta +
r\dot{\theta}\cos{\theta}].$$
The effective acceleration due to gravity is
$${\bf g}^{'}= (g - r\Omega^2\sin^2\theta)\hat{\bf r} -
r\Omega^2\sin\theta\cos\theta\hat{\bf\theta},$$

\ni where $g=GM_{BH}/(r-r_g)^2$.
Since the transverse motion of the flux tube imparts
kinetic energy to the surrounding  fluid, the effective mass
$m_i + m_e$ appears in the terms corresponding to transverse
acceleration. The drag is modeled as the drag experienced by a
cylinder moving perpendicular to its axis in a fluid, and the
drag per unit length is given by
\begin{equation}
{\bf D} = -{1\over 2}C_D \rho_e\sigma|(\dot{r}-u_{acc})\hat{\bf r} +
r\dot{\theta}\hat{\bf \theta}|\{(\dot{r}-u_{acc})\hat{\bf r} +
r\dot{\theta}\hat{\bf \theta}\},
\label{eq:drag}
\end{equation}
where $C_D = 0.4$ is a dimensionless coefficient which has a
roughly constant value of 0.4 at high Reynold's numbers
(Goldstein 1938) and $\rho$ is the density. The accreting matter is
taken to be flowing radially inward everywhere in the accretion disk and the
velocity of accretion is approximated as
\begin{equation}
u_{acc} = \alpha\sqrt{2GM_{BH}\over{(r-r_g)}},
\label{eq:u_acc}
\end{equation}
where $\alpha$ is the viscosity parameter, and is assumed to be
$0.01$ throughout the paper. Since we are interested only in the
qualitative effects which infall velocity would produce,
we use this very simple minded equation.

The flux tube is in pressure equilibrium with
the surroundings throughout its motion in the disk
\begin{equation}
p_{r,i} + p_{g,i} + {B^2\over 8\pi} = p_{r,e} + p_{g,e},
\label{eq:ipbal}
\end{equation}
where the $p_r$ and $p_g$ are the radiation and gas
pressures. The flux tubes are buoyant and the buoyancy factor is
$$
M = {(\rho_e-\rho_i)\over\rho_e}={(m_e-m_i)\over m_e},
$$
where $\rho$ is the density.

Equations (\ref{eq:ARC1})-(\ref{eq:ARC3}) are valid
for thin flux tubes which have radii small compared to the
scale heights of the ambient fluid.
Switching over to dimensionless quantities with $\xi =
r/r_g$, $\tau = (c/r_g)t$ and after dividing by $(m_i+m_e)$,
Eqns. (\ref{eq:ARC1})-(\ref{eq:ARC3}) can be written as
\begin{eqnarray}
\ddot{\xi} - \xi\dot{\theta}^2 +
{X\over{(1+X)}}[-\xi\dot{\phi}^2\sin^2\theta -
2\xi\omega\dot{\phi}\sin^2\theta] =\nonumber\\
 = {X\over{(1+X)}}\{{M\over X}[g - \xi\omega^2\sin^2\theta] -
T_{ens}\sin\theta - {D_r\over{\pi\sigma^2\rho_e}}\},
\label{eq:one}
\end{eqnarray}
\begin{eqnarray}
\xi\ddot{\theta} + 2\dot{\xi}\dot{\theta} +
{X\over{(1+X)}}[-\xi\dot{\phi}^2\sin\theta\cos\theta -
2\xi\omega\dot{\phi}\sin\theta\cos\theta] =\nonumber\\
 = - {X\over{(1+X)}}\{{M\over X}\xi\omega^2\sin\theta\cos\theta +
T_{ens}\cos\theta + {D_{\theta}\over{\pi\sigma^2\rho_e}}\},
\label{eq:two}
\end{eqnarray}
and
\begin{eqnarray}
\xi\sin\theta\ddot{\phi} + 2\dot{\xi}\sin\theta(\dot{\phi} +
\omega) + 2\xi\cos\theta\dot{\theta}(\dot{\phi} + \omega) +\nonumber\\
 + l\nott(n-2)(\xi\sin\theta)^{(n-2)}[\dot{\xi}\sin\theta +
\xi\dot{\theta}\cos\theta] = 0,
\label{eq:three}
\end{eqnarray}

\ni where $\omega=(r_g/c)\Omega$ is the dimensionless local
angular velocity of the disk. The magnetic tension is
\begin{equation}
T_{ens} = {4\pi M\nott
T_e(\xi\nott)\over{\mu_e A(1-M\nott)\xi\nott\sin\theta\nott}},
\label{eq:Tens}
\end{equation}
where subscripts $o$ refer to the initial values of the
corresponding quantities, $T$ is the temperature and
$A=(\sigma/\sigma\nott)^2$ is the expansion factor which is the
ratio of the cross-sectional area of the flux tube with its
initial cross-sectional area, which can be derived using
the mass and flux conservations equations as

\begin{equation}
A = ({T_e(\xi\nott,\theta\nott)\over{T_e(\xi,\theta)}})^3
({\xi\nott\sin\theta\nott\over{\xi\sin\theta}})
({1-M\nott\over{1-M}}).
\label{eq:Aad}
\end{equation}

\noindent {\bf 2.3 Assumptions regarding the condition inside a flux tube}

The expression $M$ in above equations strictly
depends on the nature of energy transfer processes between the disk and
the flux tube and the initial entropic condition of the tube.
We consider two extreme possibilities: (i) in $\S$ 2.3.1 when the tube
is locally isothermal with respect to the surrounding disk;  and (ii)
in $\S$ 2.3.2 when the tube moves adiabatically inside the disk. The realistic
case should lie somewhere in between these two cases; and can, in principle,
be solved by incorporating the energy transport equation. Throughout
these calculations we choose the initial condition such that
the flux tube is in thermal equilibrium with its surroundings. From
Eqn. (\ref{eq:ipbal}), the initial buoyancy factor is
$M\nott=B^2/(8\pi p_{g,e})$.

\noindent{\it 2.3.1 When the flux tubes move isothermally}

Here, flux tubes are considered to be in thermal equilibrium
with the surroundings throughout their motion in the disk;
$T_i({\xi, \theta}) = T_e({\xi, \theta})$.
Then in Eqn. (\ref{eq:ipbal}), $p_{r,e} = p_{r,i}$.
{}From the flux and mass conservation $B/\rho_i\xi\sin\theta$
should be constant. Using this and Eqn. (\ref{eq:ipbal}) we can get
a quadratic equation in $M$, whose solution is of the form

\begin{equation}
M={1\over 2}[b \pm \sqrt{b^2-4}],
\label{eq:Mai}
\end{equation}
where
$$b = 2 + ({\xi\nott\sin\theta\nott\over{\xi\sin\theta}})^2
({T_e(\xi\nott,\theta\nott)\over T_e(\xi,\theta)})^2
{(1-M\nott)^2\over M\nott}.
$$
The solution with positive sign is unphysical as it gives a
value of $M>1$ and this is possible only if the internal density is negative.

\ni{\it 2.3.2 When the flux tubes move adiabatically}

In this case there is no heat exchange between
the flux tube and its surroundings. The entropy of the flux tube
remains constant throughout its motion in the disk. In other
words $\mu_i$ and $\beta_i$ are constant and are equal to their
initial values. Flux and mass conservations demand that $B/(\rho_i
\xi\sin\theta)$ should be conserved. Using this fact in Eqn.
(\ref{eq:ipbal}) and a reasonable assumption that the flux tube
is in thermal equilibrium just before it is released,
we can derive an expression for $\rho_i/\rho_e$,

\begin{equation}
{k_1{\big (}{\rho_i\over\rho_e}{\big )}^{4/3} + k_2\big(}
{\rho_i\over\rho_e}{\big )}^2  - 1 = 0,
\label{eq:Maa}
\end{equation}

\ni where

$$ k_1 = {(1-\beta_eM\nott)\over (1-M\nott)^{4/3}},$$

$$k_2 = \beta_e{M\nott\over(1-M\nott)^2}({T_e\over T_{e,\circ}})^2
({\xi\sin\theta\over\xi\nott\sin\theta\nott})^2.$$

\ni The accretion disk being adiabatic, and $\mu$ being chosen to be a
constant, $\beta_e$ becomes contant everywhere in the disk.

\noindent{\bf  2.4 Approximate analytical behavior of the flux tubes}

Before we plunge into obtaining detailed numerical results describing the
dynamics of the flux tubes, it may be instructive to
study some approximate analytical properties of the governing equations
to obtain some insight into the problem.

First, let us write the equations (\ref{eq:ARC1}-\ref{eq:ARC3}) in
the Cartesian coordinate
centered at the black hole with the assumption $\rho_e \sim \rho_i$,
$$
2 {\ddot x} - 2 {\dot y} \Omega = M g_{eff} x - D_1 v_p {\dot x}
\eqno{(17a)}
$$
$$
2 {\ddot y} + 2 {\dot x} \Omega = M g_{eff} y - D_1 v_p {\dot y}
\eqno{(17b)}
$$
$$
2 {\ddot z}= \frac{Mgz}{\xi} - D_1 v_p {\dot z}
\eqno{(17c)}
$$
where, $D_1= C_D/2\pi\sigma$, $M$ is the magnetic buoyancy,
$g_{eff}=g/\xi-\Omega^2$; $g\sim 1/\xi^2$ is
the approximate expression for gravity, $v_p$ is the magnitude of the
poloidal velocity component.

To begin with, we consider the $z$ component of the equation.
It is rewritten as
$$
{\ddot z} +A_1 {\dot z} + A_2 z =0,
$$
where, $A_1= D_1 v_p / 2$ and $A_2=-Mg/2\xi$. In the WKB approximation
an acceptable solution is:
$$
z=C e^{mt}
$$
with $C$ a constant and $m=(-A_1+\sqrt{A_1^2-4A_2})/2$. The flux tube rises
exponentially from the place of release with a time constant of $\tau=1/m$.
In the limit of very small drag force, $D_1^2 v_p^2 << 8Mg/\xi$,
$\tau \propto {\xi^3/M}^{1/2}$. Thus, the flux tubes are eliminated
faster if the buoyancy is larger and as one approaches the rotational axis.
In the limit of a very large drag force,
$$
\tau \propto \frac{\xi^3}{M\sigma} .
$$
Thus, the time taken to expel smaller flux tubes are longer.

In solving $x$ and $y$ components of the equation, we first assume the
drag force is negligible. In the limit of
very low effective gravity (i.e., close to the center of the disk) or
very low buoyancy, the solution of Eqns. (17a-b), becomes,
$$
u_x= u_{x0} e^{i(\Omega^2 - M g_{eff})^{1/2}}.
$$
Thus, flux tubes undergo oscillations with frequency close to the local
rotational velocity.

The effect of drag is to dampen the oscillation. This can be seen
easily by assuming that the drag and buoyancy are negligible in the
$y$ direction. The solution for the velocity becomes,
$$
u_x=u_{x0} e^{-({D_1 v_p t}/{2})} e^{i\Omega^{'} t} ,
$$

\ni where $\Omega^{'}^2 = \Omega^2-Mg_{eff}/2$.
Clearly, the amplitude of velocity oscillation decreases faster for flux
tubes of smaller size.

In the presence of the two opposing forces, namely, Coriolis force and
buoyancy, the amplitude of oscillation of a flux tube released at $\xi=\xi_0$
can be calculated by simply balancing them. It becomes,
$$
\Delta \xi = - \frac{M g_{eff} (\xi_0)}{2 n \xi_0^{2n-4} l_0^2}.
\eqno{(18)}
$$
\setcounter{equation}{18}
The amplitude rises with the magnetic buoyancy and with the effective
gravity. The oscillation should commence in a direction opposite to
the sign of $g_{eff}$, namely, in the direction of the buoyancy force.
The oscillation amplitude should be higher as $n \rightarrow 0$,
namely, as the Coriolis force becomes progressively weaker.
When the Coriolis force is zero, i.e., when $n=0$, the flux tubes
shoot out in the direction of the local pressure gradient force.
This behavior of the tubes could be directly verified by ignoring
drag and tension terms and putting $d\phi/dt =0$ in equations
(\ref{eq:ARC1}-\ref{eq:ARC3}). It is to be noted that since the $z$
equation is decoupled from those of $x$ and $y$, the flux tubes
oscillate while moving in the direction parallel to the rotation
axis. This behavior has already been noted in the case of the sun
(Choudhuri \& Gilman 1987). For a given $M$ the oscillation amplitude
is larger in thick disks ($n\sim 0$) than in the sun ($n\sim 2$).

\noindent {\large {\bf 3. THE NUMERICAL RESULTS}}

In this Section, we consider a thick accretion disk, as in Figs. 1(a-b),
around a black hole of mass $10^6$ M\solar. To reduce the computational
time, we assume a small disk with the outer edge at 20 $r_g$; the
general conclusion will not depend upon the actual size of the disk,
which could be as large as $1000 r_g$.
Since we are interested only in the generic behavior of the
flux tubes and not in modeling any particular
system, we stick to these typical parameters throughout our calculations.
Using a simple boundary condition, temperature $T(R\nott) = 0\deg $K,
Eqn. (\ref{eq:addisk}) gives the structure of the
adiabatic disk for any prescribed angular momentum distribution which
is detemined by the values of $l\nott$ and $n$.
As a working definition, we
regard the chimney to be that part of the surface of the disk, which
corresponds to the region between the inner edge and the distance where
the maximum height of the disk occurs.
In \S 3.1, we first
study the behavior of flux rings in isothermal equilibrium in two
different cases: (i) with a constant
angular momentum distribution ($n=0$ and $l\nott=2$), (ii) with an
angular momentum determined by some typical values of $n=0.1$ and
$l\nott=1.75$. In \S 3.2, we study them when they move adiabatically
inside the adiabatic disk. Section 3.3 discusses the importance of
magnetic tension in our calculations.

\ni{\bf 3.1 When the flux tubes move isothermally}

The isothermal condition (See Eqn. (\ref{eq:ipbal})) restricts the value of
$M$ between $0$ and $1$ throughout its motion in the disk in order
that the density and pressure ratios remain positive.
In $\S$ 3.1.1 we present the results when both the drag and the
accretion velocity are neglible. In $\S$ 3.1.2, we introduce these effects.

\ni{\it 3.1.1 When drag and accretion flows are absent}

We place flux rings with the initial buoyancy factor $M\nott = 0.1$
at various radii $\xi=$ $4$, $5$, $6$, $7$ and $10$ in a constant angular
momentum disk ($n=0$), very close to the equator ($\theta = 89\deg$), and
release them with zero initial velocities. We assume drag and accretion
to be negligible ($D=0$, $u_{acc}=0$) and numerically integrate Eqns.
(\ref{eq:one}), (\ref{eq:two}) and (\ref{eq:three}) with $M$ taken from
Eqn. (\ref{eq:Mai}). The trajectories of the flux tubes are plotted
in the $\xi-\theta$ plane as shown in Fig. 2a. They move out along
the direction of the pressure gradient force. The tubes released
beyond the center of the disk are expelled toward the
outer edge of the disk. Only those
released between the inner edge of the disk and its center emerge
in the chimney. The markers on the curves are drawn at time intervals of
100 $r_g/c$, which gives an idea of the `buoyancy time-scales' for
the flux tubes. Flux tubes with different $M\nott$ behave identically,
except that the emerging time decreases with increasing $M\nott$ because
of the increased buoyancy. For any given $M\nott$, inclusion of drag just
increases the emerging time of flux tube without affecting its trajectory.
We may note here that flux tubes which are released strictly on the
equatorial plane of the disk remain on the plane throughout; only those
released between the inner edge and the center fall toward the black hole.

We repeat above computations for an $n=0.1$ disk, where the
angular momentum gradually rises with axial distance.
Here Coriolis force plays a very important role
in the dynamics of weak (small $M\nott$) flux tubes, but the behavior of the
strong flux tubes remains unaffected as buoyancy dominates the
Coriolis force. Figure 2b shows the results. Coriolis force dominates
in the case of $M\nott=0.01$ flux tubes; they oscillate with a small
amplitude and move almost parallel to
the rotational axis before emerging into the chimney. Flux tubes with
higher $M\nott$ ($0.1$ and $0.9$ as shown) oscillate with larger
amplitudes. Equation (18) provides an approximate expression for the
amplitude of oscillations. Note that $M\nott=0.9$ fields
behave as in the $n=0$ disk, the amplitude of oscillation being so large
that they are expelled even before they turn around.

The condition whether the flux tubes emerge into the chimney, or get expelled
out depends on the relative strengths of Coriolis force and magnetic buoyancy.
This is dictated by the angular momentum distribution (the parameters
$n$ and $l\nott$) of the disk and the field strength (governed by $M\nott$).
A simple back-of-the-envelope calculation will give the limit of
$M_{\circ,c}$ below which flux tubes released beyond the center
of a $n=0.1$ disk emerge into the chimney. We do these calculations
for flux tubes very close to the equatorial plane. The Coriolis force
experienced by a flux tube is $F_c = 2v_{\phi}\omega$, where
$v_{\phi}=\xi(\omega_1-\omega)$ is the azimuthal velocity acquired by the
flux tube when taken from a radius $\xi\nott$ to $\xi_1$ on the equatorial
plane and $\omega=l\nott(\xi\sin\theta)^{n-2}$. Since angular momentum is
conserved, $\xi_1^2\omega_1= \xi\nott^2\omega\nott$. Coriolis force turns
out to be

\begin{equation}
F_c=2l\nott^2\xi^{n-3}(\xi\nott^n-\xi^n).
\label{eq:F_CorF}
\end{equation}

\ni Magnetic buoyancy is given by

\begin{equation}
F_{MB}=M\nott[{1\over 2(\xi-1)^2} - l\nott^2\xi^{2n-3}].
\label{eq:F_MB}
\end{equation}

\ni Equating the two expressions above, we get the critical $M_{\circ,c}$
below which the flux tubes will emerge into the chimney,

\begin{equation}
M_{\circ,c}={2l\nott^2\xi^{n-3}(\xi\nott^n - \xi^n)\over
{1/2(\xi-1)^2} - l\nott^2\xi^{2n-3}}.
\label{eq:CorFMB}
\end{equation}

\ni For $n=0.1$ the chimney extends roughly from $\xi\sin\theta=3.2$
to $13.8$. A flux tube released at $\xi\nott=7$ will emerge in the
chimney $\xi<13.8$, provided $M<M_{\circ,c}=0.2$. Figure 3 shows
the trajectories of flux tubes released at $\xi\nott=$ $5$ and $7$,
for $n=0.1$ and different $M\nott$ as indicated. Flux tubes weaker than
$M\nott= 0.4$ emerge within the chimney. The discrepancy of a factor of $2$
is because, in the rough calculation, $M\nott$ is assumed to be constant
along its motion. However, as the flux ring moves out, it enters into
a region of lower pressure, and expands in order to maintain the pressure
equilibrium, which brings down its $M$ value.

Figure 4 shows two plots of $M_{\circ,c}$ with $n$ (Eqn. \ref{eq:CorFMB})
for $l\nott=2$ and $1.75$ as indicated, when $\xi\nott=7$ and $\xi=13.5$.
Flux tubes which lie in the parameter regime above the curve
are expelled outside the chimney and those below it emerge inside
the chimney. For example, for $n=0$, all flux tubes, regardless of field
strength are expelled out when released at any point beyond the center
of the disk (Fig. 2a). The sharp rise of the $l\nott=1.75$ curve at $n=0.12$
is because for $l\nott=1.75$ and $n>0.12$, the center of the disk lies
beyond $\xi\nott=7$; all flux tubes released at $\xi\nott=7$ emerge in
the chimney regardless of their
$M\nott$ value. This curve also shows that fields weaker than $M_{\circ,c}
\sim 0.25$ when released at $\xi\nott=7$ emerge inside the chimney,
which is indeed shown in the numerical result of Fig. 3.

Figure 5 shows the
trajectories of the flux tubes released at $\xi=5$ and $7$ for different
values of $n$ (indicated on the curve) and for $M\nott=0.1$ and $l\nott=2$.
In disks with $n>0.3$, Coriolis force overwhelms the buoyancy  for
$M\nott<0.1$ tubes and will make them emerge inside the chimney.
The rough calculation of the $l\nott=2$ curve in Fig. 4 predicts
this value of $n$.

\ni{\it 3.1.2 When drag and accretion flows are present}

We now investigate the effect of accretion on the behavior of these
these flux tubes. We repeat the above calculations by including
drag and accretion in the Eqns. (\ref{eq:one}) to (\ref{eq:three})
by using the equations for the drag and the accreting flow from Eqns.
(\ref{eq:drag}) and (\ref{eq:u_acc}). Figure
6a shows the trajectories for the $n=0$ disk for various values
of tube cross-sectional radii $\sigma\nott= 0.1$, $0.01$ and $0.003$ (in units
of $r_g$) as indicated. Smaller the radius, easier it is for the
flow to drag it along. Trajectories for large flux tubes,
$\sigma\nott= 0.1$, resemble the `no drag' case shown in Fig. 2a. When
the size is decreased, the flow drags the flux tubes and makes
them emerge in the chimney. Flux tubes entering the disk along with
the accreting matter can also emerge from the chimney as shown in Fig. 6b
and 6c. Here, the calculations are done by releasing the flux tubes
at the outer edge of the disk $\xi= 19$ and $\theta=85\deg$; Fig. 6b
shows the effect of the flow on tubes with varying size $\sigma\nott$
and Fig. 6c, on varying magnetic buoyancy factor $M\nott$. Figure 6b is drawn
for flux tubes with $M\nott=0.1$ and various $\sigma\nott$ as indicated. All
flux tubes with $\sigma\nott<0.005$ enter the chimney, larger tubes
are returned back towards the outer edge of the disk. Figure 6c is drawn for
flux tubes with $\sigma\nott=0.001$
and different $M\nott$ values. Fields with $M\nott<0.4$ emerge in the
chimney, whereas the stronger ones are expelled. The stronger
ones can emerge in the chimney if the sizes are
small enough. Figure 7 shows the parameter space of $M\nott$ and
$\sigma\nott$. Flux tubes below the curve get accreted into the
chimney from the outer edge of the disk and those above out do
not enter the disk at all. The $n=0.1$ curve in the Fig. 7 is
for the $n=0.1$ disk which clearly indicates the effect of Coriolis force
which pushes back the incoming flux tubes. Hence, a larger drag force, i.e.,
a smaller size, is needed to accrete the flux tubes into the chimney.

At the outer edge of the disk, magnetic buoyancy is more or less
radially outwards and opposes the drag force. The drag force due
to the flow (which is chosen to be radial with magnetude $\alpha$ times
the free fall velocity) can be approximately written as

\begin{equation}
F_D={\alpha^2C_D\over 2\pi\sigma(\xi-1)}
\label{eq:F_D}
\end{equation}

\ni In a disk, where the angular momentum varies, i.e., $n\not=0$, Coriolis
force opposes drag. The force increases as the flux tube is dragged inward.
Equating the sum of the Coriolis force [from Eqn. (\ref{eq:F_CorF})]
and magnetic buoyancy [from Eqn. (\ref{eq:F_MB})] with the drag force, we get

\begin{equation}
\sigma_{\circ,c}={\alpha^2C_D\over 2\pi(\xi-1)}\{M[{1\over 2(\xi-1)^2}
 - l\nott^2 \xi^{(2n-3)}] + 2l\nott^2\xi^{n-3}[\xi\nott^n - \xi^n]\}^{-1},
\label{eq:F_D=CorF+MB}
\end{equation}

\ni which gives the limiting value of $\sigma\nott$ that can
be accreted inside the chimney, where the values of $\xi\nott=19$
and the outer edge of the chimney is at $\xi=13.8$
for $n=0.1$ disk and $13.2$ for $n=0$ disk. The dashed curves in
Fig. 7 show the analytical curves, which gives a good resemblence
to the computed curves. The departures seen are because the analytical
calculation considers $M\nott$ instead of $M$. In the absence of
Coriolis force ($n=0$), $\sigma_{\circ,c}\propto 1/M$, all flux tubes
regardless of $\sigma$ emerge in the chimney as $M\rightarrow 0$. Since
Coriolis
force opposes drag, its presence limits the value of $\sigma$ that
can emerge into the chimney. This is evident in the Figure where
$\sigma_{\circ,c}\rightarrow 0.004$ as $M\rightarrow 0$ when $n=0.1$.
However in the large $M$ limit, this trend is reversed; flux tubes
of larger size can emerge in the chimney as $n$ increases. This is
because the effective buoyancy gets reduced as $n$ increases Eqn.
(\ref{eq:F_MB}). The cross-over seen in the analytical curves for $n=0$ and
$n=0.1$ is due to this non-linear behavior for large $M\nott$ values. The
numerical calculations also show it, though at a large value of
$M\nott\sim 0.85$.

\ni{\bf 3.2 When the flux tubes move adiabatically}

The adiabatic condition is incorporated in the calculations, by
using Eqn. (\ref{eq:Maa}) to compute $M$ when numerically integrating Eqns.
(\ref{eq:one}) to (\ref{eq:three}).
If the flux tubes move inside the adiabatic disk isentropically,
in the presence of negligibly small accretion ($u_{acc}=0$) and drag ($D=0$),
they move out of the disk due
to magnetic buoyancy alone. As they move out, they enter cooler
regions and hence expand in order to attain the pressure balance
with the surroundings. This expansion is more than it would be
in the isothermal case, because the tube is hotter now, and hence
is more buoyant. Figure 8 shows the trajectories of the flux tubes
under similar conditions as in Fig. 2a, except that they move
adiabatically. Both figures look identical, except
that the emerging times in the present case are slightly smaller than
in Fig. 2a, as the markers reveal. The rest of the dynamics is identical to the
isothermal case with an extra buoyancy force. We therefore do not study this
case in any more detail.

\ni{\bf 3.3 Importance of magnetic tension}

All the rough analytical calculations presented in the last two
sections ignore the effects of magnetic tension, which tends to bring the
flux tubes towards the rotation axis. As we shall show below, magnetic
tension is negligible for the radiation pressure dominated ($\beta<<1$)
disks that we use. However, as $\beta$ is increased, disks become
hotter, the effect of tension becomes important.

In the absence of an accreting flow, in a $n=0$ disk, magnetic
buoyancy expels the flux tubes released beyond the center of
the disk (Fig. 2a). Magnetic tension acts in the direction towards
the rotation axis. Close to the equatorial plane, magnetic tension
$F_T=B^2/(4\pi\xi\rho_e)$ and buoyancy (Eqn. \ref{eq:F_MB})
oppose each other.
When the flux tube is in isothermal equilibrium with its surroundings,
$M=B^2/(8\pi p_{g,e})$. Equating these two forces, we get

\begin{equation}
T_e={\mu c^2\over R}{\xi\over
2}[{1\over 2(\xi-1)^2} - l\nott^2\xi^{2n-3}].
\label{eq:F_TF_MB}
\end{equation}

\ni This equation gives the minimum temperature beyond which tension
dominates over buoyancy. At $\xi=7$, this temperature turns out to be
4$\times 10^{10}$ \deg K. The radiation dominated ($\beta=10^{-4}$)
adiabatic disks considered here, have typically a
central temperature of $3\times 10^6\deg$ K for $n=0$ disk and
$1.9\times 10^6\deg$ K for $n=0.1$ disk, which is several
orders of magnitude below this critical temperature. In all these
cases, magnetic tension does not play any role in the dynamics of
these flux tubes. For a larger $\beta$,
the temperature of the disk can be sufficiently high so as to exceed the
threshold for which tension dominates buoyancy and all flux tubes
emerge in the chimney.
Figure 9 shows the trajectories of $M\nott=0.1$ flux rings released
at $\xi\nott$= $5$ and $7$ for the same adiabatic disk as in Fig. 2a,
except that $\beta$ values are varied as indicated. There
is no drag force or accretion and the flux tubes rise due to buoyancy alone.
Even for $\beta=0.1$, buoyancy dominates and the trajectories resemble
those in Fig. 2a. The effect of tension becomes
pronounced when $\beta$ becomes $0.3$, and the flux tube released
at $\xi=7$ rushes towards the chimney. Indeed, Equation (\ref{eq:addisk})
shows that the temperature at $\xi=7$ is $0.86\times 10^{10}\deg$ K,
which is very close to the critical temperature estimate of
$4\times 10^{10}\deg$ K given by Eqn. (\ref{eq:F_TF_MB}).
This shows that the effect of tension is important only in a highly
gas-pressure dominated disk.

\noindent {\large{\bf 4. CONCLUDING REMARKS}}

In this paper, we have studied the dynamics of flux tubes which are
released inside or at the outer edge of an adiabatic thick accretion disk.
The general conclusions we draw are as follows. Flux tubes released between the
inner edge and the center of the disk emerge in the chimney,
regardless of the angular momentum distribution,
the magnitude of the accretion flow, size and intensity of the flux
tubes. Flux tubes released outside the center of the disk can also
emerge into the chimney depending upon the interplay among the
various disk and flux tube parameters, as well as the location
from where the tubes are released. In general, it is observed
that fields can emerge in the chimney (a) if Coriolis force
(which arises in the presence of any non-zero angular momentum gradient)
dominates over magnetic buoyancy, and (b) if the drag due to the
accreting flow dominates over magnetic buoyancy and Coriolis force.
In the presence of a large shear, as in the present case, flux tubes
tend to be toroidal giving rise to a magnetic tension force acting towards
the rotation axis. Tension dominates over all forces, aiding their
emergence into the chimney if the temperature of the disk is sufficiently
high, as could be the case for a disk surrounding a stellar mass black hole.

Chimneys are known to be the most luminous region of a thick accretion
disk, since the effective gravity is very strong. Cosmic radio
jets are believed to be originated from this region. The
jets of BL Lacs, and Optically Violent Variables are supposed to be pointing
towards us, and their variabilities are believed, but not proven,
to be caused by the magnetic activities in the chimneys. In the
present paper, we indeed see that there exists a large range of the
parameter space which renders the chimney region to be magnetically active.
We also see evidence for flux tubes being expelled away towards the outer
disks.  No flux tubes are seen to be advected {\it directly into} the
black hole, unless they are released exactly on the equatorial plane
between the inner edge and the center of the disk.

The behavior of flux tubes studied here depends upon a
number of apparently free parameters: $l_0$, $n$, $\beta$,
$u_{acc}$, $M$ and $\sigma$. Most of these parameters, however, are closely
inter-related through viscous
mechanism operating inside the disk, and at present its nature
is rather poorly understood. The deviation of $n$ from
$0$ is a measure of the viscosity. For accretion to occur, since
the total angular momentum at the inner edge must be close to the
marginally bound value, viscosity also limits the
acceptable range of $l_0$. Similarly, $\beta$, which is a measure of
radiation content of the flow, depends on the accretion rate, which is, in
turn, dictated by the viscosity. The accretion velocity is also directly
governed by the viscosity parameter $\alpha$.

The parameters, $M$ and $\sigma$, of the flux tubes that are generated
inside the disk will depend upon the details of the velocity
field, which is again decided by the nature of viscosity. The status of the
magnetic activity in the chimney will, therefore, be clearer as a
better understanding of the viscous process inside the disks
emerges. Unfortunately, there is no satisfactory discussion in the literature
on the exact nature of the viscosity in an accretion disk, though
some of the recent works make encouraging progress (e.g., Vishniac
\& Diamond, 1989; Geertsema \& Achterberg, 1992).
The parameters of the accreted flux tubes, however,
should not depend upon the viscous processes inside the disk.

We have ignored several effects,
for instance, we have considered only axisymmetric, toroidal
flux tubes. This is clearly oversimplified. In general, all the
components will be present as the field enters the disk and it is likely
to be more diffusive and filamentary in nature, rather than
in the form of axisymmetric flux tubes. However, as it enters,
differential rotation will progressively make the toroidal
field stronger. In that respect, our assumption of toroidal
tube may be justified. A diffused poloidal field, however, could
affect the dynamics of the disk (Balbus \& Hawley 1991).
It is shown that a thin, ion pressure dominated
magnetized disk is violently unstable under axisymmetric perturbation,
in the presence of even a small poloidal field. It is not clear if
this instability persists in the radiation pressure dominated
thick disks which we consider here. Furthermore, our study is restricted
to the dynamics of isolated flux tubes in a non-magnetic disk
and therefore this instability is probably irrelevant.
We also believe that the general conclusions we obtained
using axisymmetric flux tubes remain valid, even when
non-axisymmetric  flux tubes are considered
as shown by a large amount of recent work in the context of solar physics
(Morano-Insertis, 1986; Choudhuri, 1990, D'Silva \& Choudhuri, 1993).
We have also ignored amplification of the
field by shear, convection and advection. Similarly, our choice of
adiabatic equation of state in the disk is over-simplified and
nowhere inside the disk flux tubes are anchored to produce a stable
corona which might cause a sustained magnetic activity.
In our next paper (D'Silva \& Chakrabarti 1993), we present
models of accretion disks where we show that flux anchoring
by the accretion disk is possible. We also add the effects of shear
amplification. A large number of physical processes, such as magnetic activity
of BL Lacs, OVVs, the formation and collimation of jets,
acceleration of cosmic rays, etc., could be triggered by the
corona of a thick disk, detailed discussions on which are differed to
Paper II.

\ni{\it Acknowledgments:} We thank Arnab Rai Choudhuri for discussions.

\newpage

\centerline {REFERENCES}

\ni Abraham, R.G., \& McHardy, I. 1989, in Proc.\ 23rd ESLAB Symp.  on
Two Topics in X-Ray Astronomy, ed.\ J.\ Hunt \& B. Battrick
(ESA SP-296) p.\ 865\\
\ni Acheson, D. J. 1978, Phil. Trans. Roy. Soc., A289, 459\\
\ni Balbus, S.A. \& Hawley, J.F., 1991, ApJ, 376, 214\\
\ni Begelman, M., Blandford, R.D. \& Rees, M.J. 1984, Rev. Mod. Phys.,
56, 255\\
\ni Biermann, L., 1950, {\it Zs. f. Naturforsch.}, 5a, 65.\\
\ni Blandford, R.D. \& K\"onigl, A. 1979, ApJ, 232, 34\\
\ni Blandford, R.D. and Payne, D.G., 1982, MNRAS, 199, 883\\
\ni Campbell, C. G. 1992, Astrophys. \& Geophys. Fluid Dyn., 63, 197 1992\\
\ni Carini, M.T., Miller, H.R., Noble, J.C. \& Goodrich, B.D. 1992,
AJ, 104, 15\\
\ni Chakrabarti, S.K. 1985, ApJ, 288, 1\\
\ni Chakrabarti, S.K., 1986, ApJ, 303, 582\\
\ni Chakrabarti, 1990, Com. Ap., 4, 209\\
\ni Chakrabarti, S. K. 1991, MNRAS, 252, 246\\
\ni Chakrabarti, S.K. \& Bhaskaran, P. 1992, MNRAS, 255, 255\\
\ni Chakrabarti, S.K., Jin, L. \& Arnett, W.D. 1987, ApJ, 336, 572\\
\ni Chakrabarti, S.K. \& Khanna, R. 1992, MNRAS, 256, 300\\
\ni Choudhuri, A. R. 1990, ApJ, 355, 733\\
\ni Choudhuri, A. R. \& Gilman, P. A., 1987, ApJ, 316, 788\\
\ni Choudhuri, A. R. \& D'Silva, S. 1990, A\&A, 239, 326\\
\ni Coroniti, F. V., 1981, ApJ, 244, 573\\
\ni D'Silva, S. \& Chakrabarti, S.K., 1994, ApJ, (to appear)\\
\ni D'Silva, S. \& Choudhuri, A.R., 1993, A\&A, 272, 621\\
\ni Eardley, D. M. \& Lightman, A. P., 1975, ApJ, 200, 187\\
\ni Edelson, R., et al.  1991, ApJ, 372, L9\\
\ni Fishbone, L.G. \& Moncrief, V. 1976, ApJ, 207, 962\\
\ni Fukue, J. 1982, PASJ, 34, 163\\
\ni Galeev, A. A., Rosner, R. \& Vaiana, G. S., 1979, ApJ, 229, 318\\
\ni Gear, W.K. et al. 1986, ApJ, 304, 295\\
\ni Gear, W.K. et al. 1985, ApJ, 291, 511\\
\ni Geertsema, G.T. \& Achterberg, A., 1992, ApJ, 255, 427\\
\ni Gilman, P. A. 1970, ApJ, 162, 1019\\
\ni Goldstein, S. 1938, in Modern Developments in Fluid
Mechanics, (Oxford: Clarendon Press), p. 418\\
\ni Hughes, P.A., Aller, H.D., \& Aller, M.F. 1985, ApJ, 298, 301\\
\ni Jokipii, J.R., 1966, ApJ, 143, 961\\
\ni Jones, T.W. et al. 1985, ApJ, 290, 627\\
\ni Khardashev, N.S. 1962, Sov. Astr.-- AJ, 6, 317\\
\ni Kozlowski, M., Jaroszynski, M. \& Abramowicz, M.A. 1978, A\&A 63, 209\\
\ni Krichbaum, T. P., Quirrenbach, A., \& Witzel, A. 1992. in Variability
of Blazars, ed. E. Valtaoja \& M. Valtonen (Cambridge: Cambridge
Univ. Press) 331\\
\ni Lawrence, A., Watson, M.G., Pounds, K.A., \& Elvis, M. 1987, Nat, 325,
694\\
\ni Lynden-Bell, D. 1978, Phys. Scripta., 17, 185\\
\ni Madau, P. 1988, ApJ, 327, 166\\
\ni Maraschi, L., Reina, C., \& Treves, A. 1976, ApJ, 206, 295\\
\ni Marscher, A.P \& Gear, W.K. 1985, ApJ, 298, 114\\
\ni Marscher, A.P., Gear, W.K., \& Travis, J.P. 1992, in Variability of
Blazars,
\ni ed.\ E.\ Valtaoja \& M. Valtonen (Cambridge: Cambridge Univ.\ Press)
p.\ 85\\
\ni McHardy, I., \& Czerny, B.  1987, Nat, 325, 696\\
\ni Meyer, F. \& Meyer-Hofmeister, E., 1983, A \& A, 132, 143\\
\ni Miller, H.R. 1988, in  Active Galactic Nuclei, ed.\ H.R.\ Miller
\& P.J.\ Wiita (Berlin: Springer Verlag), p.\ 146\\
\ni Miller, H.R., Carini, M.T., \& Goodrich, B.D. 1989, Nat, 337, 627\\
\ni Moreno-Insertis, F., 1986, A\&A, 166, 291\\
\ni Moreno-Insertis, F., Sch\"ussler, M., \& Feriz-Mas, A., 1992, A\&A, 264,
686\\
\ni Paczy\'nski B. 1982, Astr. Ges. Mitt., 57, 27\\
\ni Paczy\'nski B. \& Wiita, P. 1980, Astron. Ap, 88, 23\\
\ni Parker, E. N., 1955, ApJ, 121, 491\\
\ni Parker, E. N., 1979, Cosmical Magnetic Fields, Oxford University Press\\
\ni Pudritz, R., 1981a, MNRAS, 195, 881\\
\ni Pudritz, R., 1981b, MNRAS, 195, 897\\
\ni Qian, S.J., Quirrenbach, A., Witzel, A., Krichbaum, T.P., Hummel, C.A.,
\ni \& Zensus, J.A. 1991, A\&A, 241, 15\\
\ni Quirrenbach, A., Witzel, A., Krichbaum, T., Hummel, G.A., Alberdi, A.,
\& Schalinski, C. 1989a, Nat, 337, 442\\
\ni Quirrenbach, A., Witzel, A., Qian, S.J., Krichbaum, T., Hummel, C.A., \&
\ni Alberdi, A. 1989b, A\&A, 226, L1\\
\ni Quirrenbach, A. et al. 1991, ApJ, 372, L71\\
\ni Rees, M.J., Begelman, M.C., Blandford, R.D, \& Phinney, E.S,
1982, Nat, 295, 17\\
\ni Rosen, A. 1990, ApJ, 359, 296\\
\ni Sakimoto, P. J. \& Coroniti, F. V., 1989, ApJ, 342, 49\\
\ni Scheuer, P.A.G., \& Readhead, A.C.S. 1979, Nat, 277, 182\\
\ni Schramkowski, G.P. \& Achterberg, A., 1993, Preprint\\
\ni Shakura, N. I. \& Sunyaev, R. A., 1973, A\&A, 24, 337\\
\ni Shibata, K, Tajima, T. \& Matsumoto, R., 1990, ApJ, 350, 295\\
\ni Soward, A. M., 1978, Astron. Nachr., 299, 25\\
\ni Spruit, H. C. \& van Ballegooijen, A. A., 1982, A\&A, 106, 58\\
\ni Stella, L. \& Rosner, R., 1984, ApJ, 277, 312 1988, ApJ, 331, 416\\
\ni Stepinski, T. F. \& Levy, E. H., 1990, ApJ, 362, 318\\
\ni Tout, C. A. \& Pringle, J. E., 1992, MNRAS, 259, 604\\
\ni Treves, A. et al. 1982, ApJ, 341, 733\\
\ni van Ballegooijen, A. A., 1982, A\&A, 113, 99\\
\ni van Ballegooijen, A. A.,1983, A\&A, 118, 275\\
\ni van Ballegooijen, A. A., \& Choudhuri, A. A., 1988, ApJ, 333, 965\\
\ni Vishniac, E.T. \& Diamond, P.H., 1989, ApJ, 347, 435\\
\ni Vishniac, E. T., Jin, L. \& Diamond, P., 1990, ApJ, 365, 648\\
\ni Wagner, S., Sanchez-Pons, F., Quirrenbach, A., \& Witzel, A. 1990, A\&A,
235, L1\\
\newpage

\centerline {Figure Captions}

\ni Fig. \ 1a --- Schematic diagram of an axisymmetric toroidal
flux tube in a geometrically thick accretion disk around a black hole.

\noindent Fig.\ 1b --- Isobaric surfaces of a thick accretion disk
around a Schwarzschild black hole. The parameters are $l\nott=2$,
$n=0$, $R\nott=20r_g$ and $\beta=10^{-4}$. The disk center is located
at $\xi=5.236$ and $\theta=0$.

\noindent Fig.\ 2(a-b) --- Trajectories of flux tubes in the
$R=\xi\sin\theta$ - $z=\xi\cos\theta$ plane, released
at $\xi= 4,\ 5,\ 6,\ 7,\ 10$ and $\theta=89\deg$, with zero initial
velocity and no drag, under isothermal conditions. The dotted line shows the
outer boundary of the disk. (a) $M\nott=0.1$ inside a $n=0$ disk, and the
markers are at time steps of $100 r_g/c$. (b) $M\nott$ values are as indicated,
the disk has $n=0.1$ and the `$+$' markers are at time intervals of $50$
and the `diamond' markers are at intervals of $100r_g/c$.

\noindent Fig.\ 3 --- Trajectories as in Fig. 2b for flux tubes
released at $\xi=5$ and $7$ for the various $M\nott$ values as indicated.

\ni Fig.\ 4 --- The analytical curves obtained from Eqn.
(\ref{eq:CorFMB}) for $\xi=13.5$, $\xi\nott=7$, $n=0.1$ and (a) $l\nott=2$ and
(b) $l\nott=1.75$ respectively indicating the parameter space of flux tubes
those are expelled to the outer edge of the disk and those
which emerge inside the chimney.

\noindent Fig.\ 5 --- Trajectories as in Fig. 2b for $M\nott=0.1$
flux tubes released at $\xi=5$ and $7$, for the various $n$
values as indicated. For all the curves $l_0=2$ is chosen.

\noindent Fig.\ 6(a-c) --- (a) Trajectories of flux tubes in the
$R=\xi\sin\theta$ - $z=\xi\cos\theta$ plane, with
$M\nott=0.1$, for the various $\sigma\nott$ values as indicated, released
at $\xi\nott= 5$, $7$, $10$, $14$ and $\theta=89\deg$, with zero initial
velocity, with drag and an accretion flow included, under isothermal
conditions. The dotted line shows the outer boundary of the disk.
(b) The conditions are same as in (a), but the $M\nott=0.1$ flux tubes are
released at $\xi\nott=19$ and $\theta\nott=85\deg$ for the
various $\sigma\nott$ as indicated. (c) The conditions are same as in (a),
but the $\sigma\nott=0.1$ flux tubes are released at $\xi\nott=19$
and $\theta\nott=85\deg$ for the various $M\nott$ as indicated.

\noindent Fig.\ 7 --- $\sigma\nott$ vs. $M\nott$ curve (solid) which separates
two distinct regions of the parameter space. Flux tubes released at $\xi=19$
with parameters below the curve will emerge from the chimney, whereas, those
released with parameters above the curve will be expelled in the outer disk.
Accretion flow is added. The exponents of the angular momentum distribution $n$
are indicated. The maximum height of the disk
is at $\xi=13.2$ for $n=0$ disk and $\xi=13.8$ for $n=0.1$ disk.

\ni Fig.\ 8 --- Trajectories of flux tubes under identical conditions
as in Fig. 2a, except that they move adiabatically. Comparison
of the markers reveals that these flux tubes have shorter emerging times.

\ni Fig.\ 9 --- Trajectories of $M\nott=0.1$ flux tubes released at
$\xi\nott=$ $5$ and $7$ in an adiabatic $n=0$ disk with no drag and no
accretion, but for different values of $\beta$ as indicated.
\end{document}